\documentclass[%
 reprint,
 amsmath,amssymb,
 aps,
]{revtex4-2}

\usepackage{graphicx}% Include figure files
\usepackage{dcolumn}% Align table columns on decimal point
\usepackage{bm}% bold math
\begin{document}

\preprint{AIP/123-QED}

\title[]{Binding Energies of Charged Particles on Dielectric Surfaces in Liquid Nitrogen}
% Force line breaks with \\
\author{A. Timsina}
 \altaffiliation[Current address: ]{Texas Instruments: SM, 501 W Shepherd Dr. Sherman, TX 75092, United States.}%Lines break automatically or can be forced with \\
\author{W. Korsch}%
 \email{Wolfgang.Korsch@uky.edu.}
\affiliation{ 
Department of Physics and Astronomy, University of Kentucky%\\This line break forced with \textbackslash\textbackslash
}%

%\author{J. He}
 %\homepage{http://www.Second.institution.edu/~Charlie.Author.}
%\affiliation{%
%Second institution and/or address%\\This line break forced% with \\
%}%

\date{\today}% It is always \today, today,

\begin{abstract}
A new approach for determining the binding energies of charged particles, such as ions and electrons, on dielectric surfaces in cryogenic liquids is introduced. The experimental technique outlined in this paper is employed to observe the buildup of charged particles on nonconductive surfaces using the electro-optic Kerr effect. The initial results of binding energy measurements on surfaces of deuterated tetraphenyl butadiene (dTPB)-coated and uncoated polymethyl methacrylate (PMMA) in liquid nitrogen are presented. Under these conditions, the ions or electrons displayed binding energies of less than 1 meV. Although these findings were obtained in liquid nitrogen, the methodology is not limited to cryogenic liquids and is applicable to a wide variety of fluids, with no essential dependence on temperature.
\end{abstract}

\maketitle

\section{Introduction and Motivation}

The interaction of low-energy charged particles with dielectric surfaces plays a central role in a wide range of physical, chemical, and technological processes. Quantifying the binding energies of ions and electrons at dielectric interfaces is therefore of broad relevance to fields including materials science, surface physics, electrochemistry, and detector technology, and has been a longstanding focus of experimental and theoretical surface science.

A variety of complementary surface-sensitive techniques have been developed to study adsorption and binding phenomena at dielectric surfaces, each providing access to distinct physical observables. Atomic force microscopy (AFM)~\cite{Giessibl2003}, particularly in force-spectroscopy and non-contact modes, enables direct measurements of interaction forces and potentials, from which adsorption energies can be inferred at the nanoscale. X-ray photoelectron spectroscopy (XPS)~\cite{GRECZYNSKI2020} probes core-level binding energies and chemical shifts, providing element-specific information on chemical states and charge transfer, albeit with careful consideration of surface charging and energy referencing. Temperature-programmed desorption (TPD)~\cite{KING1975} provides a direct means of extracting adsorption energies through analysis of thermally activated desorption kinetics. Electron energy loss spectroscopy (EELS)~\cite{Colliex2022}, while not a direct probe of adsorption energies, offers complementary information on electronic excitations and dielectric response near surfaces that can be sensitive to adsorbate--surface interactions. Together, these approaches illustrate the diverse experimental landscape available for characterizing charge binding at dielectric interfaces.

In this work, we introduce an alternative experimental approach for measuring the effective binding energies of charged particles on dielectric surfaces in liquids. The method exploits the controlled application of external electric fields to desorb weakly bound, physisorbed charges from the surface. The resulting field-induced anisotropy of the liquid is detected using electro-optical techniques, specifically through measurements of the Kerr effect. This approach provides indirect but sensitive access to the electric-field strengths required to overcome surface binding potentials. We apply this technique to study ions and electrons on deuterated tetraphenyl butadiene (dTPB)-coated and uncoated polymethyl methacrylate (PMMA) surfaces immersed in liquid nitrogen (LN$_2$), and we compare the resulting effective binding energies for the two surface types.

dTPB-coated plastic surfaces have been widely developed as wavelength-shifting layers in particle and neutron detection systems, particularly in applications where reduced hydrogen content is advantageous. dTPB efficiently converts ultraviolet and vacuum-ultraviolet photons into visible light, enabling effective optical readout of scintillation signals. While dTPB is not itself a radiation detector, the interaction of charged particles with dTPB-coated dielectric surfaces is of practical importance for detector operation and stability. Although the measurements presented here were performed in a cryogenic liquid, the underlying methodology is not intrinsically restricted to cryogenic conditions and is, in principle, applicable to non-cryogenic transparent liquids or dense gases.

The motivation for this study is closely connected to the nEDM@SNS experiment, originally proposed at the Spallation Neutron Source (SNS) at Oak Ridge National Laboratory. The goal of this experiment is to improve the sensitivity to the neutron electric dipole moment (EDM) by nearly two orders of magnitude, from the current upper limit of $1.8 \times 10^{-26}$~e$\cdot$cm (90\% C.L.)~\cite{Abel2020} to approximately $3 \times 10^{-28}$~e$\cdot$cm (90\% C.L.)~\cite{Ahmed2019}. Achieving this sensitivity requires exceptional control of systematic effects associated with strong electric fields applied within ultracold-neutron storage cells.

In the nEDM@SNS experimental concept, ultracold neutrons are stored for extended periods---typically several hundred seconds---in PMMA cells coated with deuterated polymer materials. These cells operate under strong and highly stable electric fields. Ionizing radiation arising from neutron $\beta$-decay, neutron capture in surrounding materials, and cosmic rays generates free charges within the storage volume. The separation and drift of ions and electrons in the applied field can lead to charge accumulation on dielectric boundaries, producing space-charge fields and electric-field distortions that represent a potential source of systematic uncertainty.

Electric-field instabilities in such systems arise from the formation of positively charged ion clusters and electron bubbles generated by ionization in the cryogenic medium. Due to their differing mobilities, these charges migrate toward electrodes and dielectric surfaces, where they can become trapped. Maintaining the required electric-field stability---at the percent level over typical measurement cycles of order $10^3$~s---therefore places stringent constraints on charge accumulation and removal processes~\cite{Ahmed2019,Golub1994}.

One proposed mitigation strategy involves periodically reversing the applied electric field to neutralize accumulated charges~\cite{Korsch2024}. However, the minimum electric field strength required for efficient charge removal is not well established. Effective desorption requires that the reversing electric field overcome the binding or trapping potential of ions or electrons at dielectric surfaces, including contributions from image-charge and polarization effects. Quantifying these effective binding energies is therefore essential for determining the electric-field thresholds required for controlled charge mitigation. Although the nEDM@SNS project was subsequently discontinued, similar experimental concepts are being actively considered for future facilities, such as the European Spallation Source (ESS), where control of charge accumulation at dielectric surfaces in strong electric fields will remain a critical experimental challenge.

The measurements presented in this work provide direct experimental input relevant to these questions by characterizing charge binding on representative dielectric surfaces under controlled conditions. More broadly, the technique demonstrated here offers a general framework for probing charge--surface interactions in liquid environments using electro-optical methods. Such capabilities are expected to be valuable for a range of applied systems in which electric-field stability and charge control at dielectric interfaces are critical.

This paper is organized as follows. Section~II describes the experimental setup and measurement principle. Section~III outlines the experimental procedures and analysis methods. The results are presented in Section~IV, followed by conclusions and outlook in Section~V.

\section{Description of Experimental Setup}

The experimental apparatus was designed to enable sensitive electro-optical measurements in a cryogenic liquid under applied electric fields. It combines a stabilized laser source, polarization optics for state preparation and analysis, a cryogenic system housing the dielectric sample, and phase-sensitive detection electronics for signal readout. The key elements of the setup are described below.
\subsection{\label{sec:level1}Cryostat}
%Fig.~\ref{fig:cryostst}%
\begin{figure}
\vspace{18pt}
\includegraphics[width=\columnwidth]{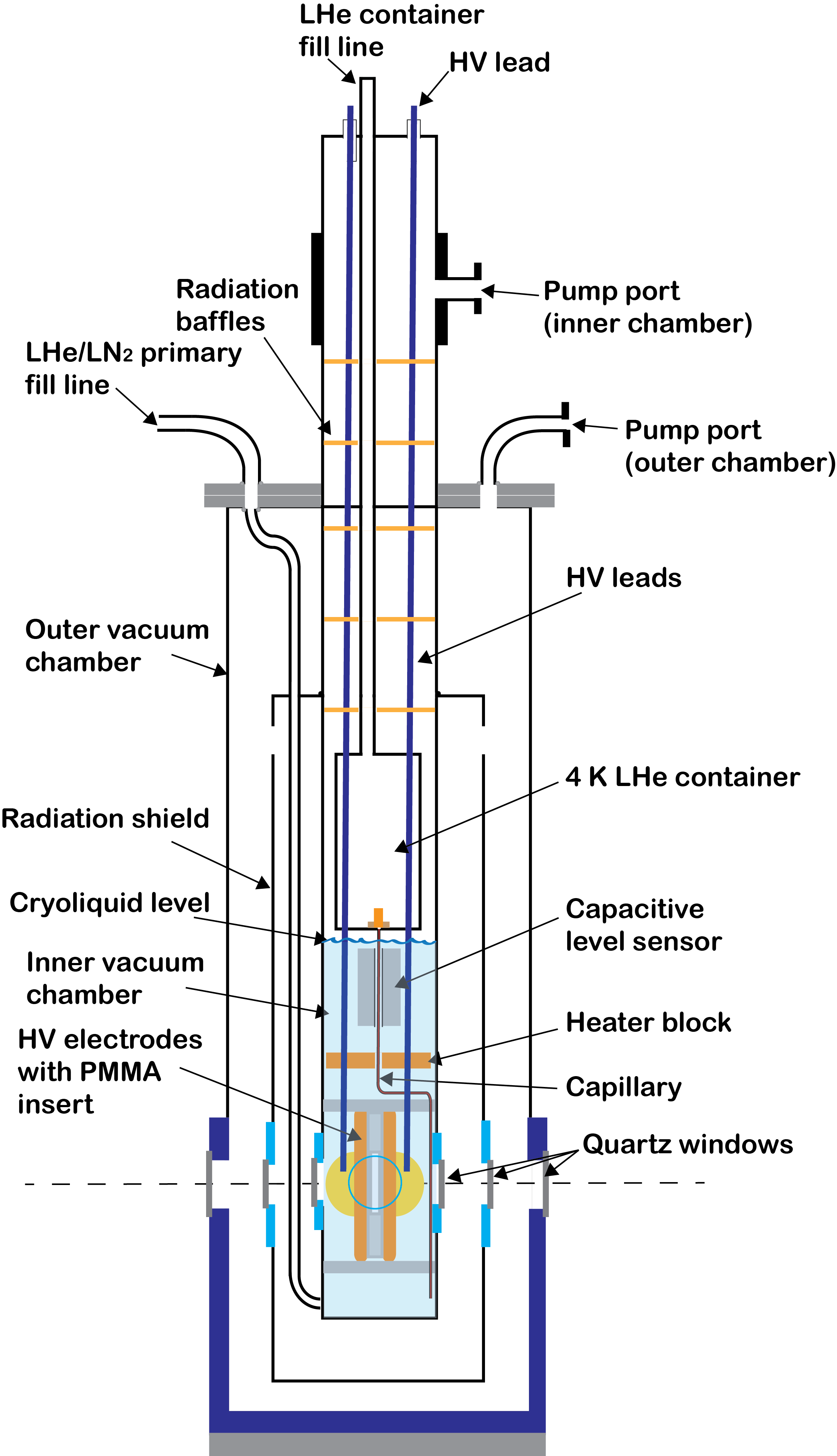}% Here is how to import EPS art
\caption{\label{fig:cryostat} Schematic of the cryostat used to measure ion binding energies. The central region of the cryostat is shown, including the electrode assembly and two PMMA inserts with a narrow optical access gap for the laser beam. Note that the LHe container and the capillary were not used in the experiments described in this paper. The figure is adapted from Fig. 6 of Ref.~\onlinecite{Korsch2024}}.\label{fig:SetupPhoto}
\end{figure}
%Fig.~\ref{fig:cryostst}%
\begin{figure}
\vspace{18pt}
\includegraphics[width=\columnwidth]{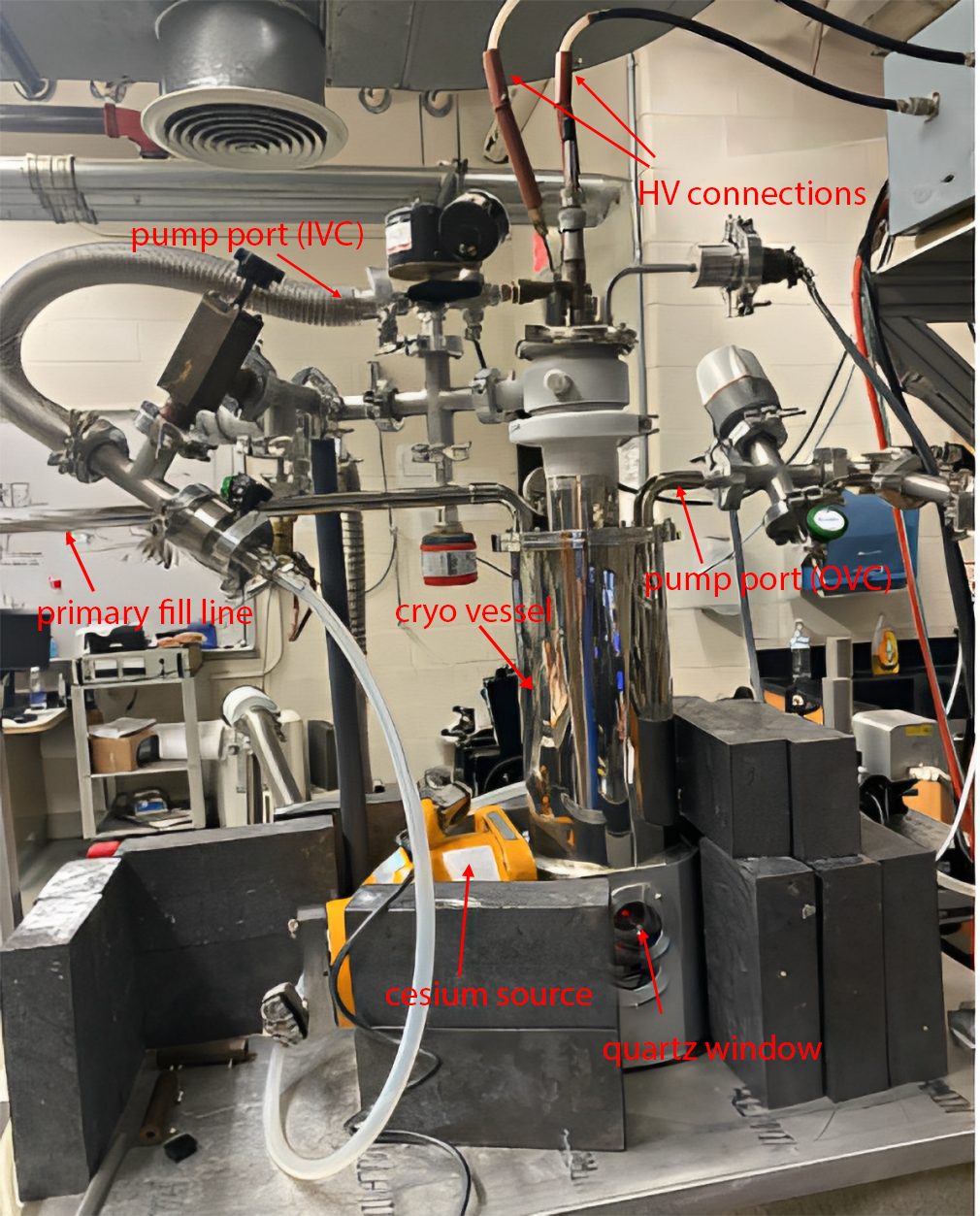}% Here is how to import EPS art
\caption{Photograph of the experimental cryostat. Visible components include the high-voltage leads, liquid-helium fill lines, pump ports, cesium source, lead shielding blocks, and optical access through quartz windows.}\label{fig:SetupPhoto}
\end{figure}
The central component of the experiment was a modified SuperTran 380 (STVP-100) continuous-flow cryostat manufactured by Janis Research. The cryogenic system was originally designed for cooling small samples to 4.2~K using continuous helium gas flow. For the present experiment, the cryostat was modified to enable continuous-flow operation with superfluid helium. A schematic of the modified system is shown in Fig.~\ref{fig:cryostat}.

The cryostat consists of an outer vacuum chamber (OVC), which thermally isolates the central system from the surrounding environment, and a compact inner vacuum chamber (IVC) that can be cooled to superfluid helium-4 temperatures ($\gtrsim 1.7$~K). An additional radiation shield located between the OVC and IVC, together with radiation baffles above the IVC, reduces the radiative heat load on the central experimental region. The high-voltage electrode assembly is located in the lower section of the IVC.

Optical access to the central region is provided by three quartz windows on each side of the cryostat. These windows are mounted in a stress-minimized configuration designed to limit mechanical strain during cool-down and warm-up. This design is essential, as temperature-dependent changes in window birefringence or optical path length would otherwise necessitate repeated calibrations at the operating temperature.

For the studies reported here, the inner central chamber containing the electrodes and dielectric materials was cooled exclusively using liquid nitrogen. Under these operating conditions, the outer vacuum chamber (OVC) reached pressures of approximately $10^{-6}$ mbar once the inner vacuum chamber (IVC) stabilized at temperatures between $\sim$60 K and $\sim$70 K. To minimize conductive heat loads, quartz windows, low–thermal-conductivity structural materials, and a dedicated mechanical support structure were employed.
The cooling system lowers the temperature of the electrode–PMMA assembly and, in the final stage, immerses the assembly in liquid nitrogen. Filling is performed exclusively through the primary fill line supplying the bottom region of the IVC. Prior to introducing liquid nitrogen, the vacuum jacket is evacuated, and both the OVC and IVC are cleaned and pumped using two independent pumping systems. To prevent solidification of liquid nitrogen within the fill line, the pressure in the IVC is maintained above 94 torr, typically at $\sim$400 mbar.
The IVC is cooled at a controlled rate of approximately 6 K/min to mitigate thermal shock. Continuous nitrogen transfer is maintained until the desired liquid level is reached and the system temperature stabilizes near 70 K. Data are acquired while the temperature remains between 70 K and 77 K, corresponding to a stable operating period of approximately 4 hours. Once the temperature apporaches the boiling point, increased light scattering of emerging bubbles degrades the laser signal and limit further data collection.
To mitigate the risk of high-voltage breakdown associated with decreasing liquid-nitrogen levels, the cryogen height is monitored using a capacitive level sensor in combination with a capacitance meter. Variations in the measured capacitance provide a continuous indication of the liquid-nitrogen level, allowing the applied high voltage to be adjusted manually before electrode exposure and electrical breakdown occur.
Figure~\ref{fig:SetupPhoto} shows a photograph of the cryogenic setup, providing a representative view of the cryostat geometry, scale, and surrounding infrastructure used in the experiment.
 
\subsection{\label{sec:level2}Optical Setup}
Figure~3 shows the optical layout used for ellipsometric analysis of the polarized probe beam. The system is designed to measure polarization changes with micro-radian sensitivity. A low-power, intensity-stabilized He–Ne laser (632.8 nm, <5 mW; Thorlabs HRS015) provides the probe beam. Beam alignment and focusing are achieved using two steering mirrors (M$_1$, M$_2$) and two convex lenses (L$_1$, L$_2$), which collimate the beam to a diameter below 1 mm in order to avoid clipping during propagation through the narrow gap between the PMMA plates. Alignment is verified by measuring the transmitted power as a function of the horizontal propagation angle and adjusting the beam position to maximize transmission while minimizing contact with the dielectric surfaces.
\begin{figure} \vspace{18pt} 
\includegraphics[width=\columnwidth]{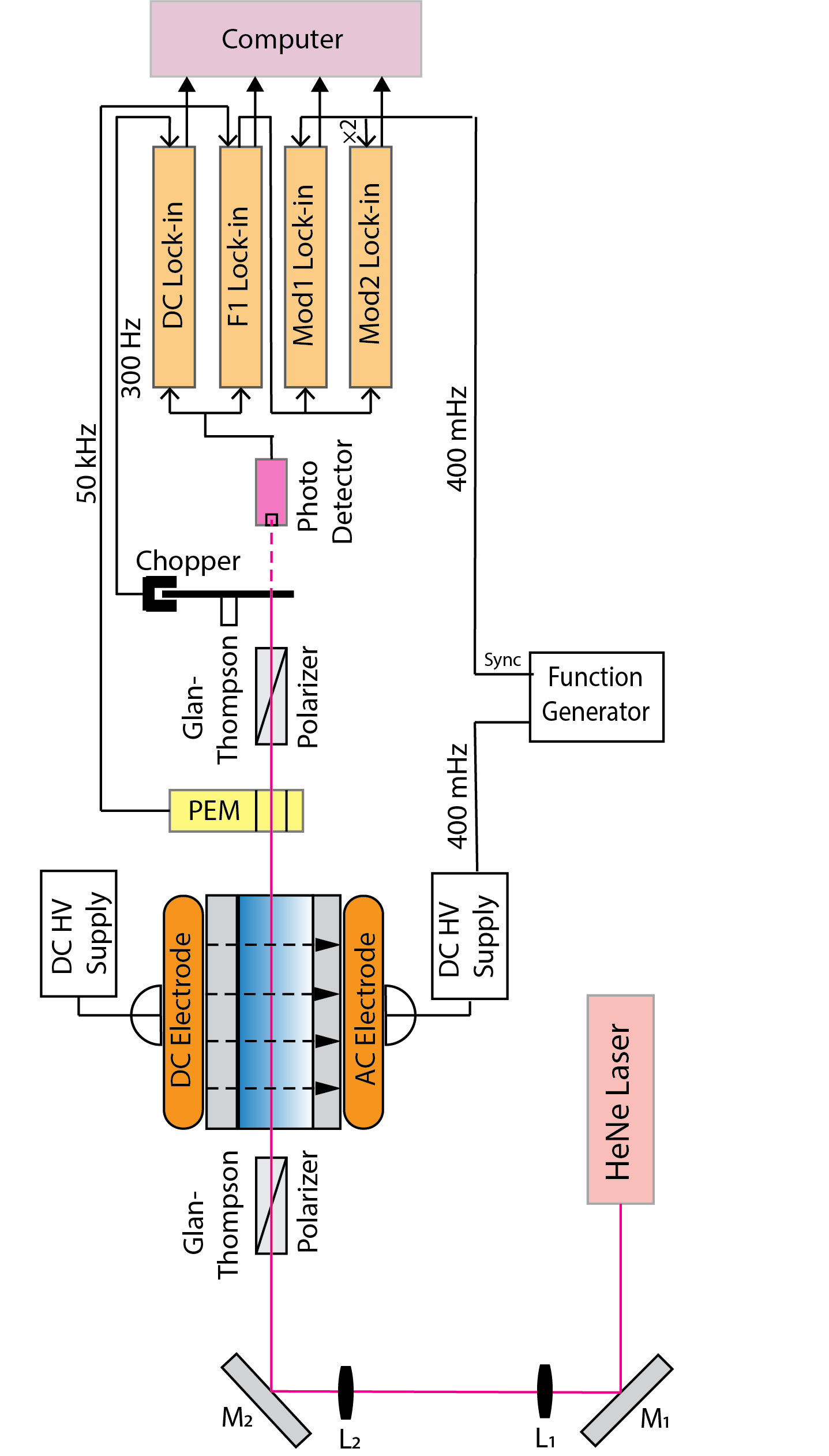}
\caption{The optics setup for the measurement of binding energy of ions and electrons on dielectric surfaces. The figure is adapted from Fig. 7 of Ref.~\onlinecite{Korsch2024}.}\label{fig:opticalsetup} 
\end{figure}
A Glan–Thompson polarizer defines the input linear polarization before the beam enters the cryostat. In this geometry, the liquid nitrogen inside the cryostat is subjected to a horizontal electric field. After transmission through the cryostat, the polarization state is analyzed using a photoelastic modulator (PEM-100, Hinds Instruments) followed by a second Glan–Thompson analyzer. The transmitted intensity is detected using a photodiode (DET-100-002, Hinds Instruments).
To achieve maximum sensitivity in the ellipsometric measurements, the incident beam is linearly polarized at an angle $\theta_p = \pi/4$ relative to the applied electric field. The photoelastic modulator is oriented at $\theta_{\mathrm{PEM}} = \pi/4$, and the analyzer prism is set to $\theta_a = 0$. This configuration provides optimal sensitivity to small polarization rotations under the present experimental conditions.
The polarization state of the transmitted light is independently characterized using a broadband polarimeter (Thorlabs PAX1000IR1, 600–1080 nm), which is used to calibrate and verify the ellipsometric measurements.
\subsection{\label{sec:level3}High Voltage Measurement}
\begin{figure}[h]
\centering
\hspace*{-0.5cm}
\includegraphics[width=\columnwidth]{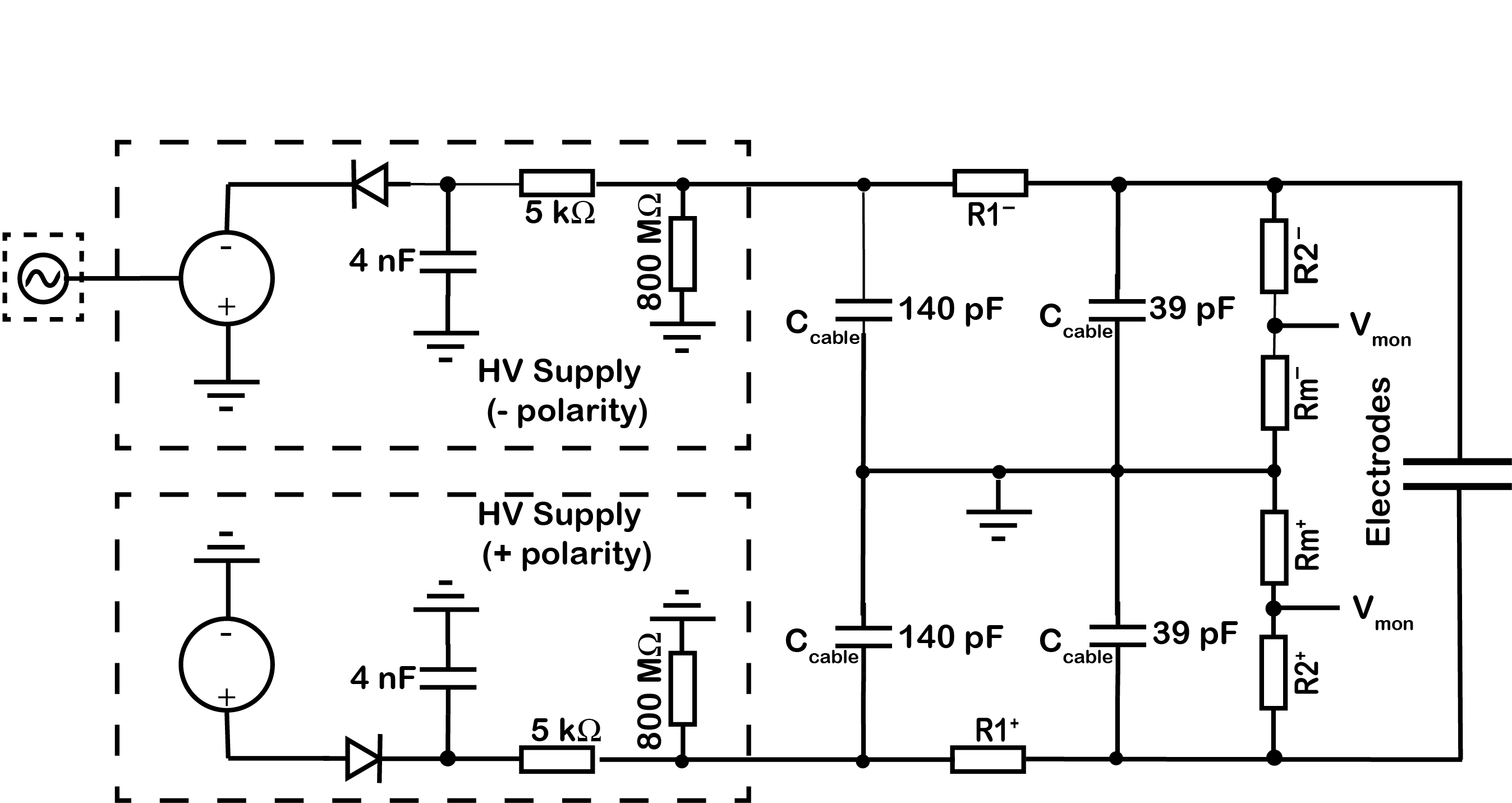}
\caption{A block diagram of the electric circuitry connected to the HV electrodes. See text for details. The figure is adapted from Fig.~8 of Ref.~\onlinecite{Korsch2024}.}
\label{fig:Resistors}
\end{figure}

% \begin{figure}
% \vspace{18pt}
% \includegraphics[scale=0.12]{figures/SL-150.jpeg}% Here is how to import EPS art
% \caption{\label{fig:HV} At the bottom left of the picture are two Spellman SL30PN150 high voltage (HV) power sources. The top items on the top shelf are a waveform generator and a digital multimeter.}
% \end{figure}
The electric field was generated using two Spellman SL30PN150 high-voltage power supplies, each connected to a separate electrode. The supplies were operated with opposite bias, providing up to 30 kV per electrode and a maximum electrode-to-electrode potential difference of 60 kV. The units share a common ground via their rear-panel ground connections. According to the manufacturer’s specifications, the supplies exhibit 0.005\% load regulation, 0.005\% line regulation, and 0.1\% peak-to-peak ripple. The negatively biased electrode was driven with a 400 mHz harmonic modulation generated by an Agilent 33120A waveform generator \cite{Broering2020}, while the positively biased electrode was held at a fixed DC potential. In both circuits, identical custom series resistance networks were used to limit current and protect the power supplies.
Both electrodes were powered using independent high-voltage circuits of identical topology, differing only in the resistor values used for current limiting. A simplified block diagram of the circuit is shown in Fig.~\ref{fig:Resistors}.
Each branch of the circuit, labeled HV$^{-}$ and HV$^{+}$ in Fig.~\ref{fig:Resistors}, comprised three resistors housed in a grounded resistance box. The measured resistance values for HV$^{-}$ were R$_{1}^{-}=$(19.93 $\pm$ 0.01) M$\Omega$, R$_{2}^{-} = $ (99.73 $\pm$ 0.01) M$\Omega$, and R$_{m}^{-} =$ 10 k$\Omega$ 
(top right in Fig.~\ref{fig:Resistors})
and the values for HV$^{+}$ were R$_{1}^{+}=$(20.96 $\pm$ 0.01) M$\Omega$, R$_{2}^{+} = $ (100.37 $\pm$ 0.01) M$\Omega$, and R$_{m}^{+}=$ 10 k$\Omega$ (bottom right in Fig.~\ref{fig:Resistors}).
All values were determined using an Agilent 3458A 8½-digit digital multimeter. The resistance values were chosen to limit the maximum possible current to less than 1.5 mA at 30 kV in case of a high-voltage break-down.  The negative electrode was driven by an AC voltage with a DC offset supplied via a function generator. 
For this configuration, the electrode voltage is given by
\begin{equation}\label{eq:VE}
V_E^{\pm} = {{R_2^{\pm}+ R_m^{\pm}} \over {R_1^{\pm} + R_2^{\pm} + R_m^{\pm}}} V_{supply}^{\pm}.    
\end{equation}
A voltage drop of 2.5 V across a 10 k$\Omega$ resistor corresponds to a 30 kV voltage at a single electrode. This voltage drop can be measured using an ADC connected to the data acquisition computer or a regular voltmeter. The voltage on each electrode is determined by multiplying the measured values by $\approx$10,000 to obtain the actual voltage on the electrode.

\subsubsection{RC Effect}\label{sec:rceffect}
The finite response of the high-voltage branch is influenced not only by the external resistors but also by the internal capacitances of the Spellman supplies ($C \approx$ 4 nF per unit). The HV$^-$ polarity was modulated at $f_{mod} = 400$ mHz ($T$ = 2.5 s). With an effective resistance of $R_{eff} \approx 800$ M$\Omega \parallel$ 120 M$\Omega \approx$104 M$\Omega$, the corresponding time constant is $\tau =$ R$_{eff}$C $\approx$ 0.42 s; this is not negligible compared with $T$. Small cable capacitances were neglected in this estimate.  The lock-in amplifiers, Mod 1 and Mod 2 in Fig.~\ref{fig:opticalsetup},
measure signals proportional to the reference frequency, $f_{mod}$, and any signal distortion causes a reduction in the base frequency amplitude by shifting some signal intensity into higher harmonics. These shifts have to be taken into account when evaluating the electric fields.
\begin{figure}
\includegraphics[width=\columnwidth]{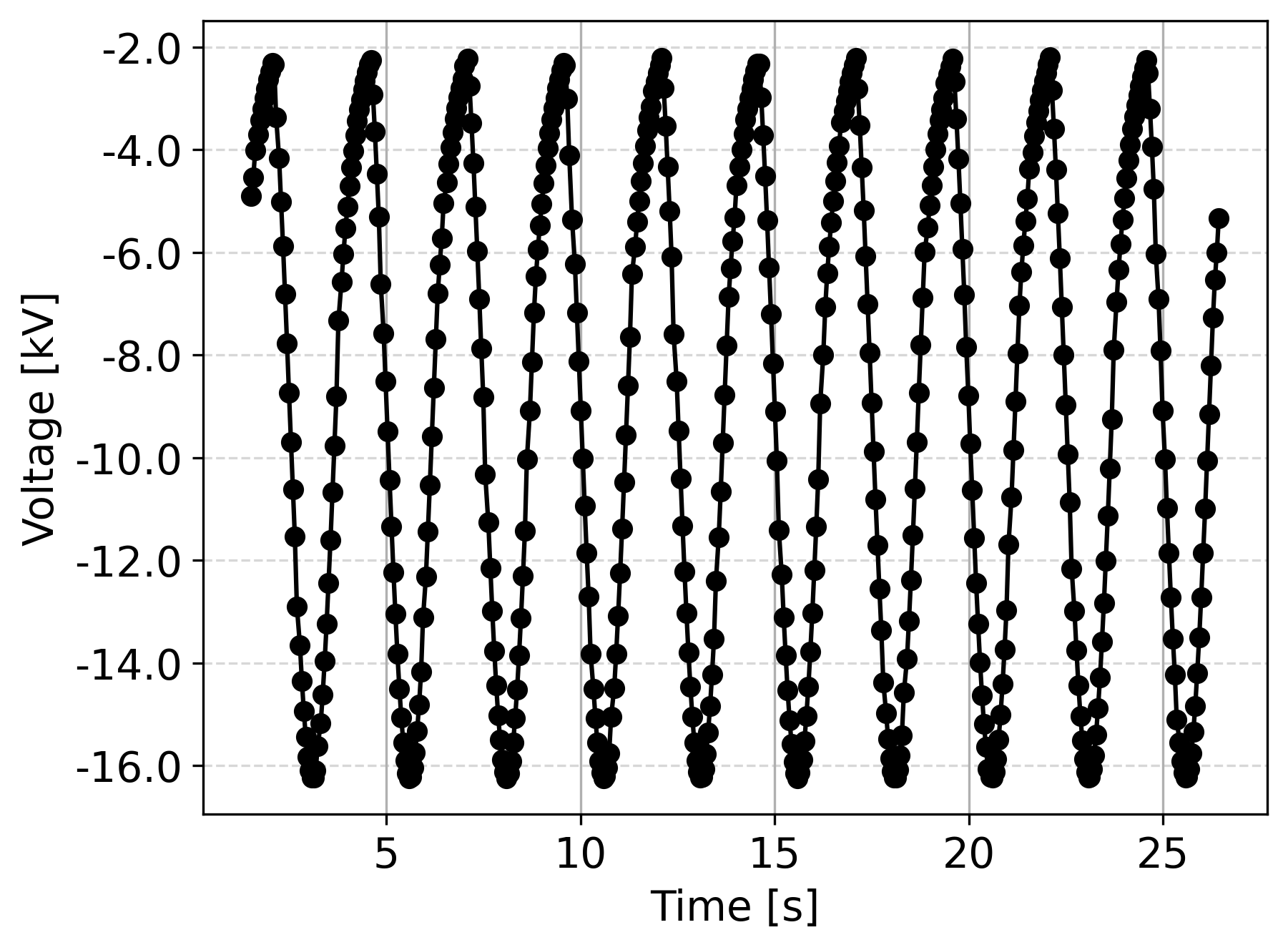}% Here is how to import EPS art
\caption{Voltage drop signal measured across the measurement resistor, $R_m^{-}$. A sinusoidal driving waveform of frequency of 400 mHz was applied to the HV${-}$ supply.}
\label{fig:VoltageSignal}
\end{figure}
\begin{figure}
\includegraphics[width=\columnwidth]{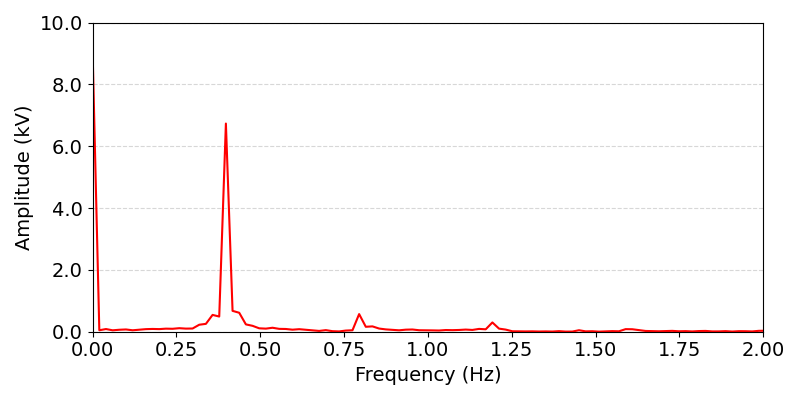}% Here is how to import EPS art
\caption{Fourier transforms of the monitored signal from FIG.~\ref{fig:VoltageSignal}. Frequency contributions from the 2$^{nd}$ (0.8 Hz), 3$^{rd}$ (1.2 Hz), and even 4$^th$ (1.6 Hz) harmonics are clearly visible.}
\label{fig:FFTVoltage}
\end{figure}
A representative AC voltage signal measured in the negative electrode circuit (HV$^{-}$) is shown in Fig.~\ref{fig:VoltageSignal}, and its Fourier spectrum is presented in Fig.~\ref{fig:FFTVoltage}. The DC component of the spectrum corresponds to a DC offset of $-$9.00 kV on the negative electrode. 

The positive electrode (HV$^{+}$) was measured as described above and yielded a DC offset of $+$5.80 kV. Therefore the  net electrode-to-electrode DC potential difference was approximately 14.80 kV. The dominant AC component appears at the modulation frequency $f_{mod} = 400$ mHz, with an amplitude of $-$6.87 kV. Given these values, the magnitude of the high voltage values oscillated between $\approx$7.93 kV and $\approx$21.67 kV.

Additional harmonics at 0.8 Hz, 1.2 Hz, and 1.6 Hz are attributed to the finite $RC$ response of the high-voltage circuit. These higher-order components account for approximately 13.3\% of the total signal amplitude. The effect of the $RC$ response was independently verified by comparing the Fourier-domain analysis with direct lock-in amplifier measurements. The RMS voltage measured by the lock-in corresponded to 86.7\% of the peak voltage across the resistor, consistent with a signal attenuation of $(13.3 \pm 0.54)$\% at $f_{mod}$=400 mHz. Circuit simulations performed using LTspice show measurable signal distortions at frequencies below approximately 10 mHz. Although operating at lower modulation frequencies would further suppress $RC$-induced attenuation, this would increase 1/$f$ noise and extend data acquisition times. The chosen operating parameters therefore represent a compromise between minimizing $RC$ losses and maintaining acceptable signal-to-noise performance, and the resulting attenuation was accounted for in the data analysis.

\section{Experimental Method}
\subsection{Electo-optic Kerr Effect}
The experimental setup was designed and constructed to satisfy several key requirements. It needed to generate electric fields on the order of several tens of kV/cm within a cryogenic liquid, such as liquid nitrogen or superfluid helium-4, and to accommodate a dummy PMMA "cell" positioned between the high-voltage electrodes in order to study the behavior and binding energies of ions on plastic surfaces. In addition, the stability of the electric field had to be continuously monitored in the presence of ionizing radiation.

To meet these requirements, the electro-optic Kerr effect was employed as a noninvasive diagnostic of the electric field. In this effect, initially linearly polarized light acquires ellipticity due to electric-field-induced changes in the refractive indices of the medium along two orthogonal directions transverse to the direction of light propagation. Although an isotropic medium does not exhibit birefringence in the absence of an external field, the application of an electric field perpendicular to the optical axis induces anisotropy, resulting in different refractive indices parallel and perpendicular to the field direction. The magnitude of this induced birefringence can be expressed as follows:
\begin{equation} \label{delta_n} 
\Delta n= n_{\parallel}-n_{\perp}=\lambda K  E^2.
 \end{equation}
In this equation, $\lambda$ represents the wavelength of the (laser) light, $K$ denotes the Kerr constant, which depends on the material, and $E$ corresponds to the magnitude of the electric field. When a linearly polarized laser beam propagates along the $z$ direction, it acquires a small ellipticity given by:
\begin{equation} \label{eq:epsilon} 
\epsilon= \frac{\pi}{\lambda}K E^2\sin(2\theta_p)L~,
 \end{equation}
here $\theta_p$ is the inital polarization angle relative to the direction of the electric field and $L$ represents the effective length of the medium that is exposed to the electric field. In this experiment, in order to maximize the sensitivity, $\theta_p$ is set to $\pi$/4. 
\par
After passing through the active medium in the cryostat, the polarization state of the light is analyzed using a photoelastic modulator (PEM). The ellipticity and rotation components of the polarization can be detected at different harmonics of the PEM modulation frequency. In particular, the ellipticity signal is extracted from the first harmonic at 50 kHz.

A schematic of the optical setup used in this experiment is shown in Fig.~\ref{fig:opticalsetup}. To ensure accurate measurements, the laser beam was additionally modulated using an optical chopper operating at approximately 300 Hz. The modulated signal was detected with a separate lock-in amplifier, labeled the “DC Lock-in” in the figure, which was configured with the same time constant as the lock-in amplifier used for the ``Mod1 Lock-in'' signal. This arrangement ensures that both the DC and Mod1 signals are integrated over identical time intervals, allowing their ratio to be formed without introducing systematic effects associated with differing temporal averaging or laser-intensity fluctuations.
\par
For the anticipated small signal strengths, the applied electric field in this study is subjected to modulation as well. As motivated in Sec.~\ref{sec:rceffect}, a low frequency of approximately 400 mHz is preferred. The time-varying electric field consists of both a direct current (DC) component and an alternating current (AC) component. Mathematically, the electric field is represented as
\begin{equation} \label{eq:efieldmod} 
E(t)= \frac{V_{DC}+V_{AC}\cos(\omega_m t)}{d}.
\end{equation}
Here, $V_{DC}$ represents the DC offset, $V_{AC}$ is the amplitude of the modulated high voltage, $\omega_m$ denotes the angular frequency, and $d$ represents the separation between the electrodes. Consequently, by plugging Eq.~\ref{eq:efieldmod} into the expression for the induced Kerr effect (Eq.~\ref{eq:epsilon}), we obtain:
\begin{equation}
\label{eq:epsvolt}
\begin{aligned}
\epsilon(t)
&= \frac{\pi K L}{\lambda d^{2}}
\Bigg[
V_{DC}^{2}
+ \frac{V_{AC}^{2}}{2}
+ 2 V_{DC} V_{AC} \cos(\omega_m t) \\
&\qquad\qquad
+ \frac{V_{AC}^{2}}{2} \cos(2\omega_m t)
\Bigg] .
\end{aligned}
\end{equation}
Equation~\ref{eq:epsvolt} naturally separates into first- and second-harmonic contributions at frequencies $\omega$ and $2\omega$, respectively. For a quantitative description of these terms, the interaction length $L$ is defined as the total distance along the optical path over which a nonzero electric field is present. In addition to the nominal electrode length, this effective length, denoted $L_{eff}$ below, includes contributions from electric-field fringe regions near the electrode boundaries, where the field extends beyond the physical electrode gap. These fringe fields increase the accumulated Kerr-induced ellipticity and must therefore be included in $L_{eff}$. The electric field inside the liquid nitrogen is not given simply by $V/d$ due to the presence of the PMMA layers. Instead, the effective
electric field in LN$_2$ is
\begin{equation}
\label{eq:Eeff}
E_{{LN_2}} =
\frac{V_{{eff}}}
{d + 2t {{\kappa_{LN_2}} \over {\kappa_{PMMA}}}}~,
\end{equation}
in this equation $d$ is the separation between the PMMA plates and $t$ is the PMMA plate thickness.
With these corrections, the first- and second-harmonic contributions to the induced ellipticity can be written explicitly as
\begin{equation}
\label{eq:eps1}
\epsilon_1(t)
= \frac{\pi}{\lambda} K L_{{eff}}
\bigl[ 2 E_{{LN_2}}^{(DC)} E_{LN_2}^{(AC)} \bigr]
\cos(\omega t) ,
\end{equation}
and
\begin{equation}
\label{eq:eps2}
\epsilon_2(t)
= \frac{\pi}{\lambda} K L_{{eff}}
\frac{\bigl(E_{{LN_2}}^{(AC)}\bigr)^2}{2}
\cos(2\omega t) .
\end{equation}

\par
To isolate the individual contributions of the induced Kerr effect, the experimental setup employed two lock-in amplifiers, labeled “Mod1 Lock-in” and “Mod2 Lock-in” in Fig.~\ref{fig:opticalsetup}, which were used to detect different harmonic components of the signal. The first-harmonic cross term, $\epsilon_1(t)$, proportional to the product $V_{DC}V_{AC}$, is sensitive to variations in the DC electric-field component and was therefore used to detect potential wall charging caused by ionization processes. Such charging would manifest as a gradual reduction of the effective DC field over time.
The second-harmonic term, $\epsilon_2(t)$, which scales with $V_{AC}^2$, provides a measure of the stability of the applied AC amplitude. However, its sensitivity is a factor of four lower than that of the first-harmonic cross term, resulting in a correspondingly reduced signal-to-noise ratio. Since the primary quantity of interest is the first-harmonic response, the second-harmonic signal was mainly used as a diagnostic to monitor the stability of the AC drive during data acquisition.
\par
Experimentally, the measured ellipticity $\epsilon_1$ is directly proportional to the ratio of the signals obtained from the ``Mod1 Lock-in" and the ``DC Lock-in" amplifiers. 
\begin{equation} \label{eq:eps_lockin}
\epsilon_1^{max} = {\pi \over \lambda} K L_{eff} \bigl[ 2 E_{{LN_2}}^{(DC)} E_{LN_2}^{(AC)} \bigr] \propto \frac{V_{Mod1}^{LIA}}{V_{DC}^{LIA}}.
\end{equation}
This ratio provides a quantification of the induced Kerr effect and allows for the determination of the impact of the applied electric field on the polarization state of the light \cite{Timsina2023}.
\par
It is important to note that the input signals for the “Mod1” and “Mod2” lock-in amplifiers were derived from a single detector output by routing the signal through the “F1 Lock-in” amplifier. To ensure accurate signal processing without distortion, the integration time constant of the “F1 Lock-in” amplifier had to be chosen appropriately. This requirement was satisfied by selecting a sufficiently short integration time that did not cause any measureable signal distortion. Given that the PEM (F1) frequency was 50 kHz and the modulation frequencies for Mod1 and Mod2 were 400 mHz and 800 mHz, respectively, this condition was readily fulfilled. Typical time constants were 20 µs for the “F1 Lock-in” amplifier and 50 s for the "Mod1” and “Mod2” lock-in amplifiers during measurements of ion and electron binding energies on the PMMA surface. Numerical simulations of the experimental setup confirmed that the influence of the chosen time constants on the measured signals was negligible.

\subsection{Experimental Procedure}\label{sec:procedure}
The binding energy measurement of ions or electrons on dielectric surfaces in cryogenic liquids presents a novel and compelling approach, offering valuable insights of the bonding nature of the charged particles within cryogenic systems. In this study, the binding energies on dTPB-coated and uncoated PMMA surfaces was systematically evaluated when they are embedded in liquid nitrogen. Detailed experimental procedures are described below. 
\begin{figure}
%\vspace{18pt}
\hspace*{-15pt}
\includegraphics[width=\columnwidth]{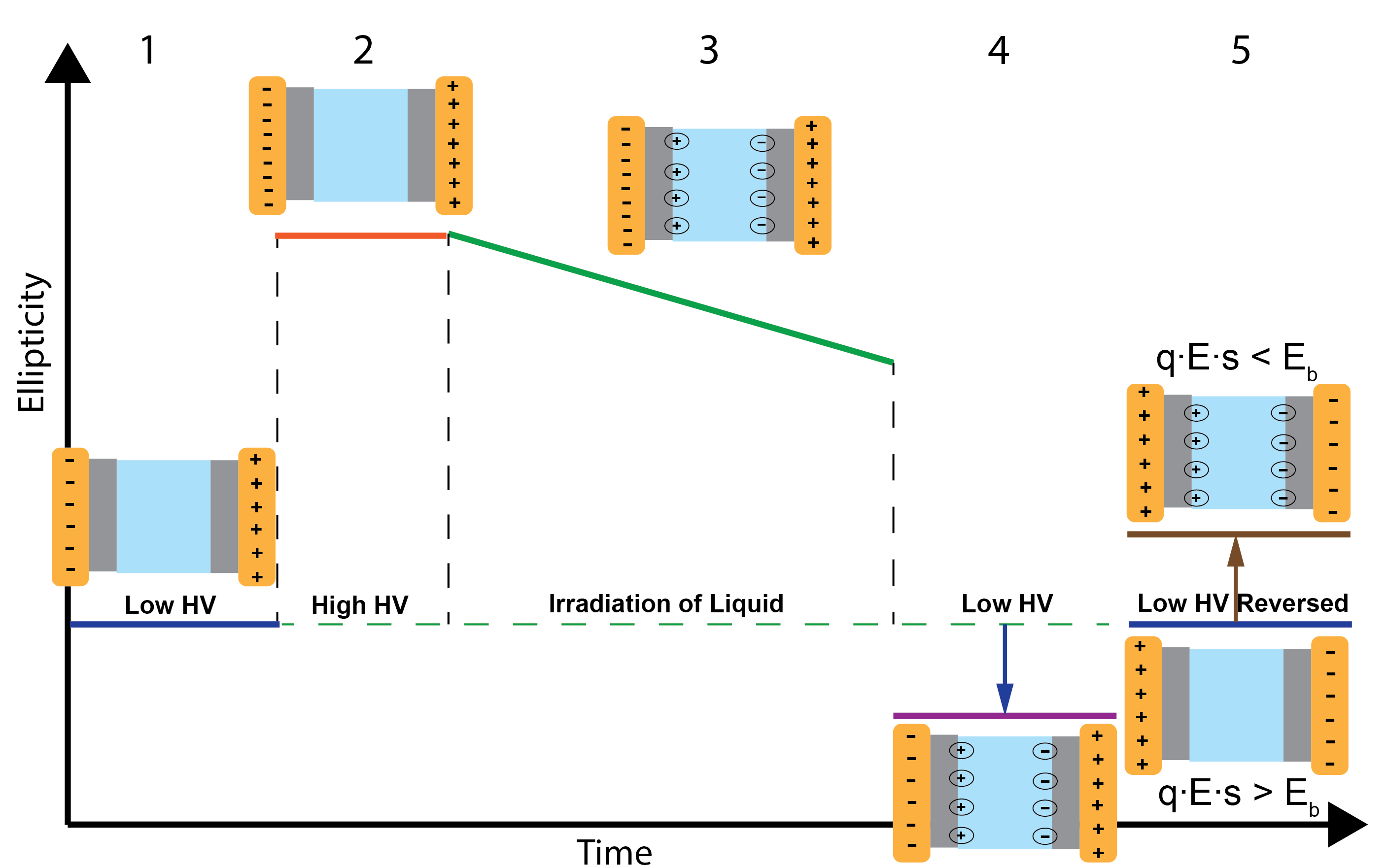}% Here is how to import EPS art
\caption{Experimental procedure for the measurement of the binding energy of ions or electrons on dielectric surfaces in liquids or dense gases.}\label{fig:BEMethod}
\end{figure}
The experimental procedure is summarized schematically in Fig.~\ref{fig:BEMethod}. Each measurement begins with the establishment of a baseline ellipticity. In this initial stage (blue line, step~1), relatively low voltages are applied, corresponding to an effective voltage of $V_{{eff}}=\sqrt{V_{{AC}}V_{{DC}}}\approx 1$–$2$~kV and, consequently, low electric fields. The resulting ellipticity defines the baseline signal, shown as the first green trace in the ellipticity–time curve of Fig.~\ref{fig:BEMethod}.
Next, the applied voltage is increased to a significantly higher value, $V_{\mathrm{eff}}\approx 11$~kV (orange line, step~2), producing a corresponding increase in ellipticity. This voltage was chosen as the maximum value that could be applied without inducing electrical breakdown. At this point, the shutter of the $^{137}$Cs source is opened, initiating ionization. The $^{137}$Cs source emits 662~keV $\gamma$ rays and 514~keV $\beta$ particles (endpoint energy 1.176~MeV); however, due to the materials located between the source and the liquid nitrogen volume, ionization is dominated by $\gamma$ radiation~\cite{Korsch2024}. The elevated electric field enhances the ionization current by increasing the probability that charges escape immediate recombination~\cite{Seidel2014}, thereby improving the charge collection efficiency.
The resulting positive nitrogen ions drift toward the negatively biased electrode and become adsorbed on the PMMA surface, while electrons migrate toward the PMMA surface near the positively biased electrode. As surface charge accumulates, the electric field generated by the adsorbed charges partially opposes the applied field, leading to a gradual reduction in ellipticity. This behavior is observed as the approximately linear decrease indicated by step~3 (green line) in Fig.~\ref{fig:BEMethod}.
After approximately one hour of charge collection, the $^{137}$Cs source shutter is closed and the applied voltage is reduced to the initial baseline value (purple line, step~4). The total collected charge remains at least two orders of magnitude smaller than the number of adsorption sites corresponding to a full monolayer on the dielectric surface~\cite{Korsch2024}. Nevertheless, the presence of surface charges results in an ellipticity signal that lies below the original baseline, as indicated by the blue trace in Fig.~\ref{fig:BEMethod}.
In the final stage of the measurement, the polarity of the low voltage is reversed (step~5). Two distinct outcomes are observed. If the reversed field is insufficient to overcome the binding energy of the adsorbed ions and electrons, the charges remain bound to the PMMA surface, and the ellipticity signal exceeds the original baseline (red trace). Conversely, if the reversed field is strong enough to induce desorption and recombination, the ellipticity signal returns to the baseline level (green trace).
The binding-energy electric field is determined by first identifying the maximum electric field for which no desorption occurs and subsequently performing separate measurement runs with incrementally increasing reversed fields. The threshold field is defined as the value at which the ellipticity signal coincides with the baseline signal, indicating complete charge removal from the surface.

\section{Measurements and Results}
\subsection{Binding Energy Measurements on dTPB-Coated PMMA Surfaces}
he binding energies of ions and electrons on dTPB-coated PMMA immersed in liquid nitrogen were measured using the experimental method described above. The system was calibrated by extracting the Kerr constant of liquid nitrogen, as described in Ref.~\cite{Korsch2024}. The value determined in that measurement was
\[
K_{{LN_2}} = (4.38 \pm 0.10) \times 10^{-18}\,\mathrm{(cm/V)^2}~,
\]
which is in excellent agreement with the value reported in Ref.~\cite{Sushkov2004}.

To ensure adequate signal-to-noise, a range of low voltages was initially tested such that the measured ellipticity exceeded $10~\mu\mathrm{rad}$. For ellipticities below this threshold, substantially longer integration times were required, which reduced the number of data points obtainable within a single liquid-nitrogen fill (typically $\sim$5~h). Based on these considerations, the lowest voltages used in the initial measurement were
$V_{{AC}}=(967 \pm 10)$~V and $V_{{DC}}=(1601 \pm 10)$~V, yielding an ellipticity of $(14.6 \pm 0.5)~\mu\mathrm{rad}$. This signal level was sufficient for the measurements presented below. A lock-in amplifier integration time constant of 50~s was chosen as an optimal compromise between statistical uncertainty and acquisition time.

The parameters entering Eq.~\ref{eq:Eeff} are $d = 2.79$ mm, $t = 0.63$ mm, $\kappa_{PMMA} = 1.91$, $\kappa_{LN_2} = 1.45$, corresponding to an effective separation $d_{eff} \approx 3.75$ mm. Using these values, the effective electric field at the center of the PMMA electrode system was calculated to be $3.32$~kV/cm.

\begin{figure}
\includegraphics[width=\columnwidth]{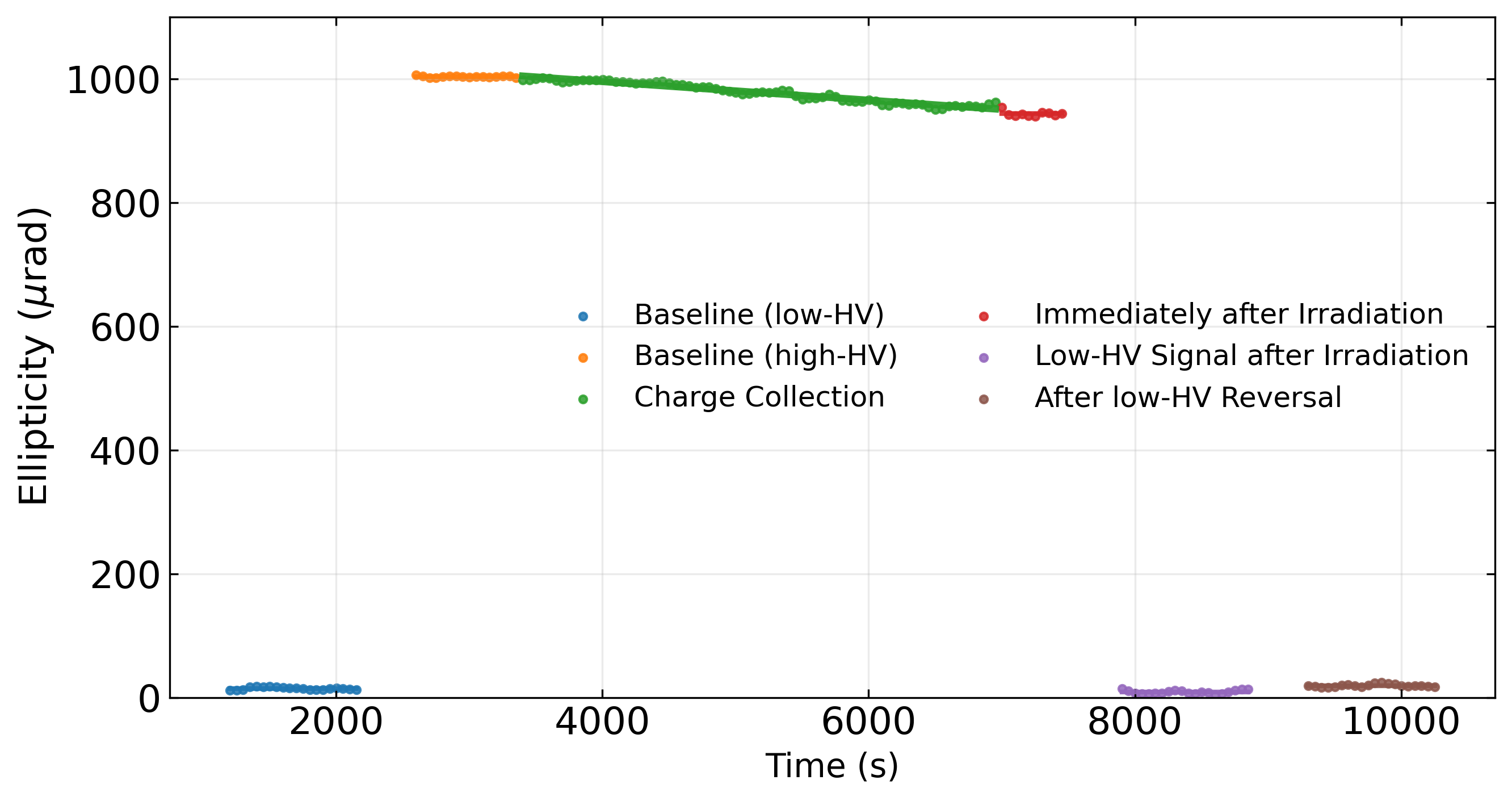}
\caption{Experimental measurement of the binding energy of ions from a dTPB-coated PMMA surface at an effective electric field of $3.32$~kV/cm.}
\label{fig:BEunzoomed}
\end{figure}

Figure~\ref{fig:BEunzoomed} shows a complete measurement cycle as described in Sec.~\ref{sec:procedure}. At an effective electric field of $3.32$~kV/cm, an initial ellipticity of $(14.6 \pm 0.5)~\mu\mathrm{rad}$ was observed (blue data points). The signal was subsequently increased to $(957.6 \pm 0.3)~\mu\mathrm{rad}$ (orange data points) by raising the voltages to $V_{{AC}}=(6871 \pm 10)$~V and $V_{{DC}}=(14801 \pm 10)$~V.

\begin{figure}
\vspace{18pt}
\includegraphics[width=\columnwidth]{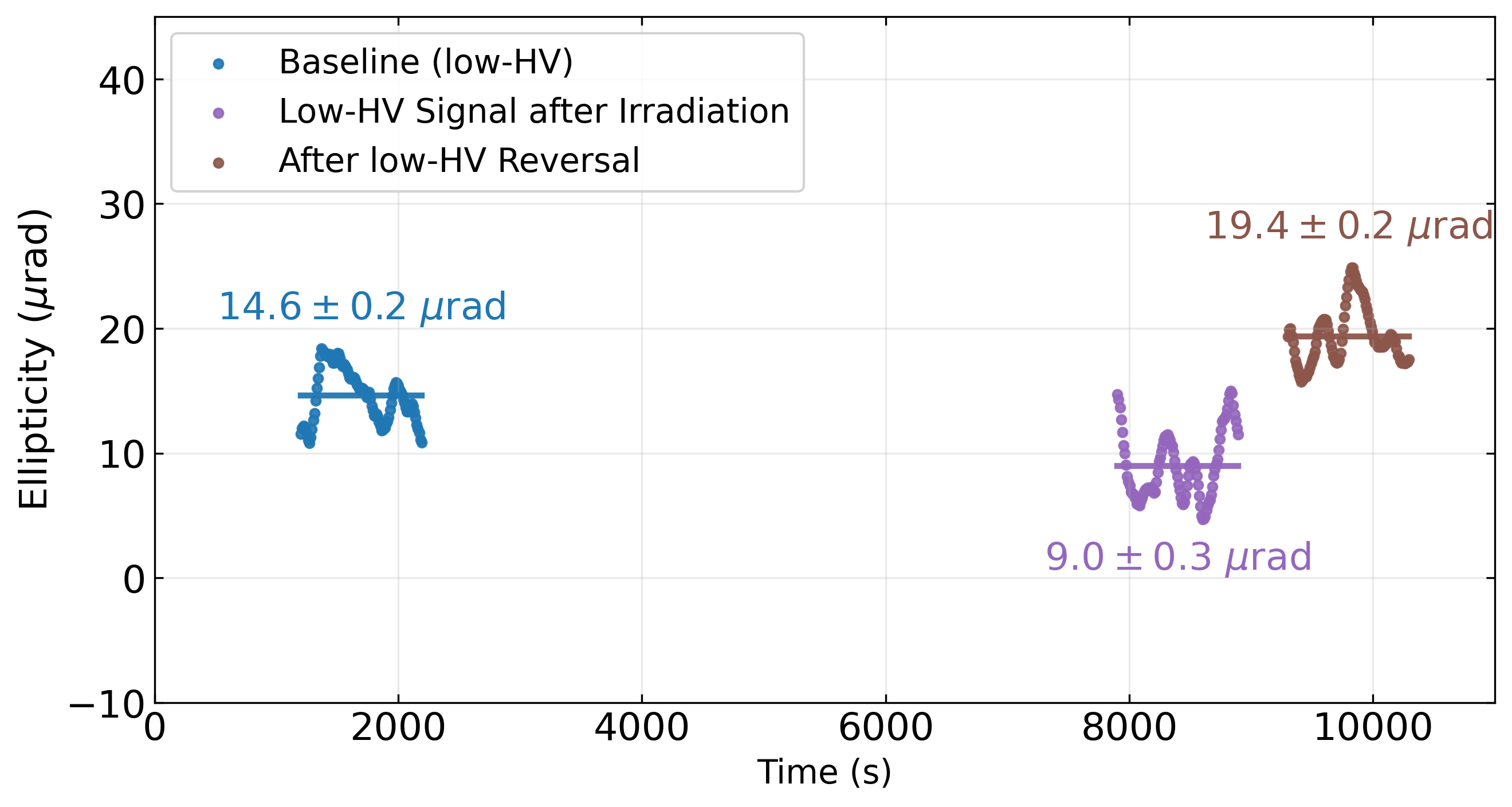}
\caption{Zoomed-in view of the low-ellipticity region shown in Fig.~\ref{fig:BEunzoomed}.}
\label{fig:BEzoomed}
\end{figure}

Following this step, the liquid nitrogen was irradiated by opening the shutter of the cesium source, resulting in a linearly decreasing ellipticity signal (green data points). This decrease indicates the accumulation of charge on the dielectric surfaces. After approximately one hour of irradiation, the voltage was reduced to the initial low-field value used to establish the baseline ellipticity. The measured ellipticity at this stage was $(9.0 \pm 0.7)~\mu\mathrm{rad}$ (purple data points), which is significantly below the original baseline, confirming that charged particles had separated and adsorbed onto the PMMA surfaces.

Reversing the low-voltage polarity resulted in an ellipticity of $(19.4 \pm 0.5)~\mu\mathrm{rad}$ (brown data points), exceeding the baseline signal. This observation suggests that the reversed electric field was insufficient to desorb the ions or electrons from the PMMA surface. For clarity, a zoomed-in view of the low-signal region is shown in Fig.~\ref{fig:BEzoomed}.

Multiple measurements were conducted following a similar procedure as previously described. These measurements involved obtaining baseline and reverse ellipticity readings at different effective electric fields. The results of these measurements are presented in Table~\ref{tab:coated}, which summarizes the measured baseline and reverse ellipticity values at various effective electric fields. 
\begin{table}
\caption{Measured baseline and reversed-voltage ellipticity values and uncertainties at different effective electric fields for dTPB-coated PMMA walls.}\label{tab:coated}
\begin{ruledtabular}
\begin{tabular}{cccccccc}
$E_{eff}$& $\epsilon_{baseline}$ & $\epsilon_{reverse}$ & $ \Delta \epsilon$ & $\delta \epsilon_{stat}$ & $ \delta \epsilon_{sys}$ & $\delta \epsilon_{tot}$ \\
\hline
 3.32 & $14.6 \pm 0.5$  & $19.4 \pm 0.5$ & 4.8 & 0.7& 0.4 &0.8\\
 3.55 & $17.3 \pm 1.0$ & $20.7 \pm 0.9$ & 3.5 & 1.3& 0.4&1.4\\
 3.80 & $20.6 \pm 0.6$& $25.0 \pm 0.3$ & 4.4 & 0.7 & 0.5&0.8\\
 4.05 & $22.3 \pm 0.9$& $24.5 \pm 1.2$ & 2.2 & 1.5 & 0.6&1.6\\
 4.31 & $24.8 \pm 0.7$ & $25.5 \pm 1.0$ & 0.7 & 1.2& 0.6&1.4\\
 4.74 & $30.6 \pm 1.2$& $30.9 \pm 0.4$ & 0.3 & 1.3 & 0.8&1.5\\
5.22 & $36.0 \pm 0.9$& $36.9 \pm 0.6$ & 0.9 & 1.1 &  0.9&1.4\\
\end{tabular}
\end{ruledtabular}
\end{table}
The table also includes the differences between the reverse and baseline signals ($\Delta \epsilon = \epsilon_{\text{reverse}} - \epsilon_{\text{baseline}}$), as well as the corresponding statistical ($\delta \epsilon_{\text{stat}}$), systematic ($\delta \epsilon_{\text{sys}}$), and total ($\delta \epsilon_{\text{tot}}$) uncertainties.
\begin{figure}
\vspace*{-10pt}
\includegraphics[width=\columnwidth]{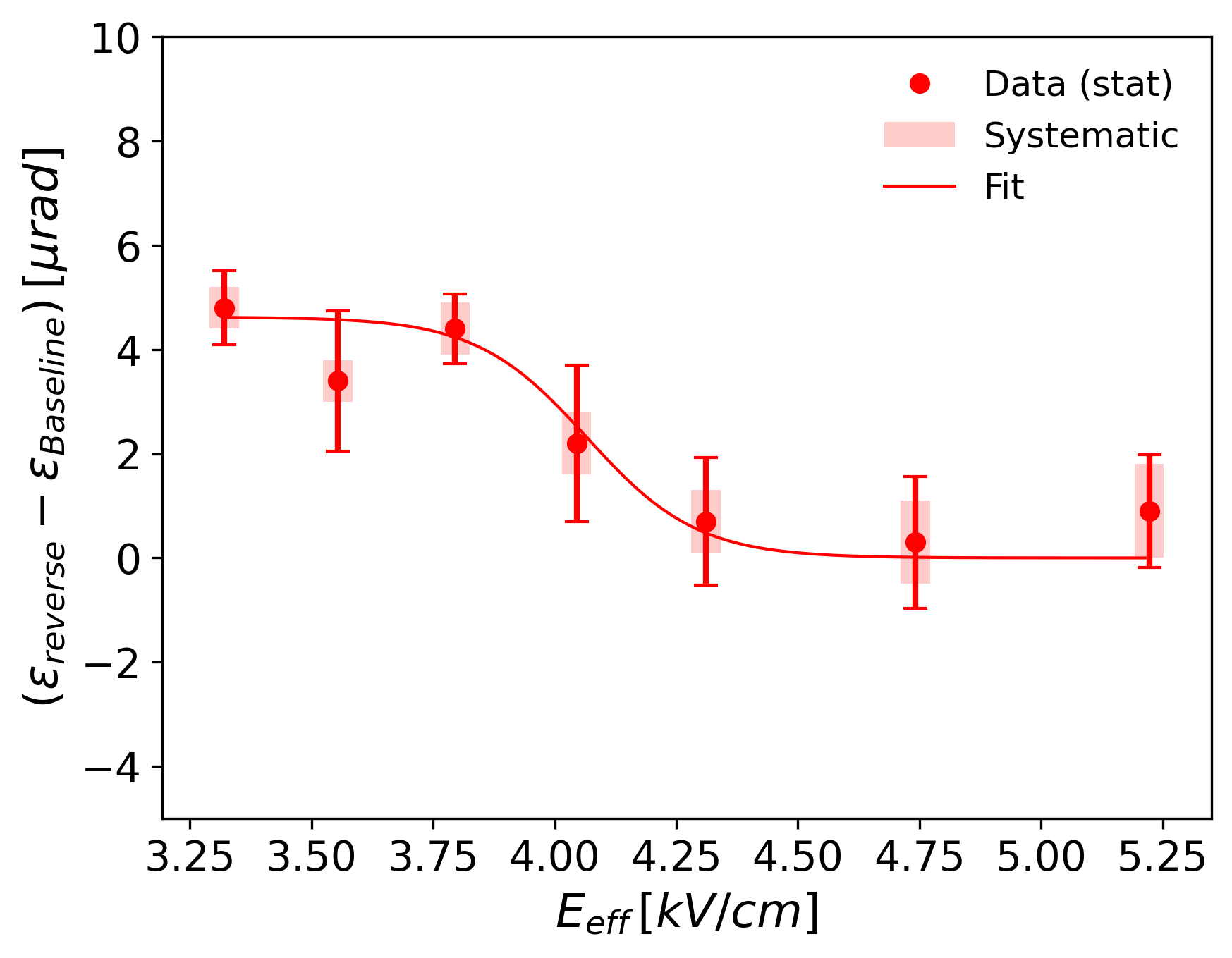}% Here is how to import EPS art
\caption{\label{fig:BECoated} Electric field dependence of the difference in reversed-voltage and baseline ellipticity measurements for dTPB-coated PMMA surfaces. Shown are the statistical (red bars) and systematic (red boxes) uncertainties with a Fermi-type function fit.}\label{fig:pmmacoated}
\end{figure}
Figure~\ref{fig:pmmacoated} shows the results of the experimental investigation of the dependence of the difference between the reversed-voltage and baseline ellipticity measurements on the effective electric field. The horizontal axis represents the effective electric field $E_{\mathrm{eff}}$ in kV/cm, while the vertical axis shows the ellipticity difference $\Delta\epsilon$ in $\mu$rad. The measurements were performed on a dTPB-coated PMMA walls immersed in liquid nitrogen using the experimental procedure described in Sec.~III(B).

The red markers indicate the measured data points, with statistical uncertainties shown as red error bars and systematic uncertainties represented by red boxes. The data were fit using a nonlinear least-squares algorithm with a Fermi-type function characterized by a cutoff electric field $E_0$ and a width parameter $\Delta E$:
\begin{equation}\label{eq:fermifcn}
\Delta \epsilon_{\mathrm{fit}} =
\frac{\epsilon_0}{1 + \exp\!\left(\frac{E_{\mathrm{eff}} - E_0}{\Delta E}\right)} .
\end{equation}
Here, $\epsilon_0$ is a scaling factor with units of $\mu$rad.

The fit results for the coated PMMA surface, obtained using the three free parameters $\epsilon_0$, $E_0$, and $\Delta E$, are shown in Fig.~\ref{fig:pmmacoated}. The fitted values are
$\epsilon_0 = (4.58 \pm 0.73)~\mu$rad,
$E_0 = (4.07 \pm 0.15)$~kV/cm,
and $\Delta E = (0.12 \pm 0.11)$~kV/cm.
The fit was performed using the total uncertainties, with statistical and systematic errors added in quadrature.

When only the statistical uncertainties are included in the fit, the resulting parameters are
$\epsilon_0 = (4.62 \pm 0.64)~\mu$rad,
$E_0 = (4.07 \pm 0.13)$~kV/cm,
and $\Delta E = (0.11 \pm 0.10)$~kV/cm.
The close agreement between the two sets of fit results indicates that systematic uncertainties have a negligible effect on the extracted parameters.

Figure~\ref{fig:pmmacoated} shows a pronounced change in the ellipticity difference, $\Delta\epsilon$, from
$(4.20 \pm 0.82)~\mu$rad at low electric fields to $(0.63 \pm 0.91)~\mu$rad at high electric fields. This change corresponds to a $2.91\,\sigma$ shift when the total uncertainty is taken into account.

The width of the transition, defined as the electric-field interval between the points where $\Delta\epsilon$ reaches 90\% and 10\% of its maximum value, is $0.51$~kV/cm. Using the effective electric field at the midpoint of the transition, $E_{\mathrm{eff}} = 4.07$~kV/cm, we estimate the strength of the interaction between the ions or electrons and the surface. This value serves as a reference field for the onset of charge removal from the PMMA surface and corresponds to a force of approximately $0.04$~meV/\AA\ acting on ions or electrons adsorbed on the dTPB-coated PMMA surface.

Based on a theoretical estimate assuming van der Waals interactions~\cite{Timsina2023}, complete removal of ions from the surface requires a separation distance of approximately $3$~\AA. Using this separation distance, the corresponding binding energy is estimated to be $0.12$~meV.

\subsection{Binding Energy Measurementson Uncoated PMMA Surfaces}
When the dTPB-coated PMMA walls were replaced with geometrically identical uncoated PMMA walls, the measurement of the binding energy was conducted using the same methodology. The measured values of baseline and reversed-voltage ellipticity at various effective electric fields are presented in Table~\ref{tab:uncoated}. As in the previous section the table also includes the differences between the reversed-voltage and baseline signals ($\Delta \epsilon = \epsilon_{\text{reverse}} - \epsilon_{\text{baseline}}$), as well as the corresponding statistical ($\delta \epsilon_{\text{stat}}$), systematic ($\delta \epsilon_{\text{sys}}$), and overall ($\delta \epsilon_{\text{tot}}$) uncertainties.
\begin{figure}
\vspace{-10pt}
\includegraphics[width=\columnwidth]{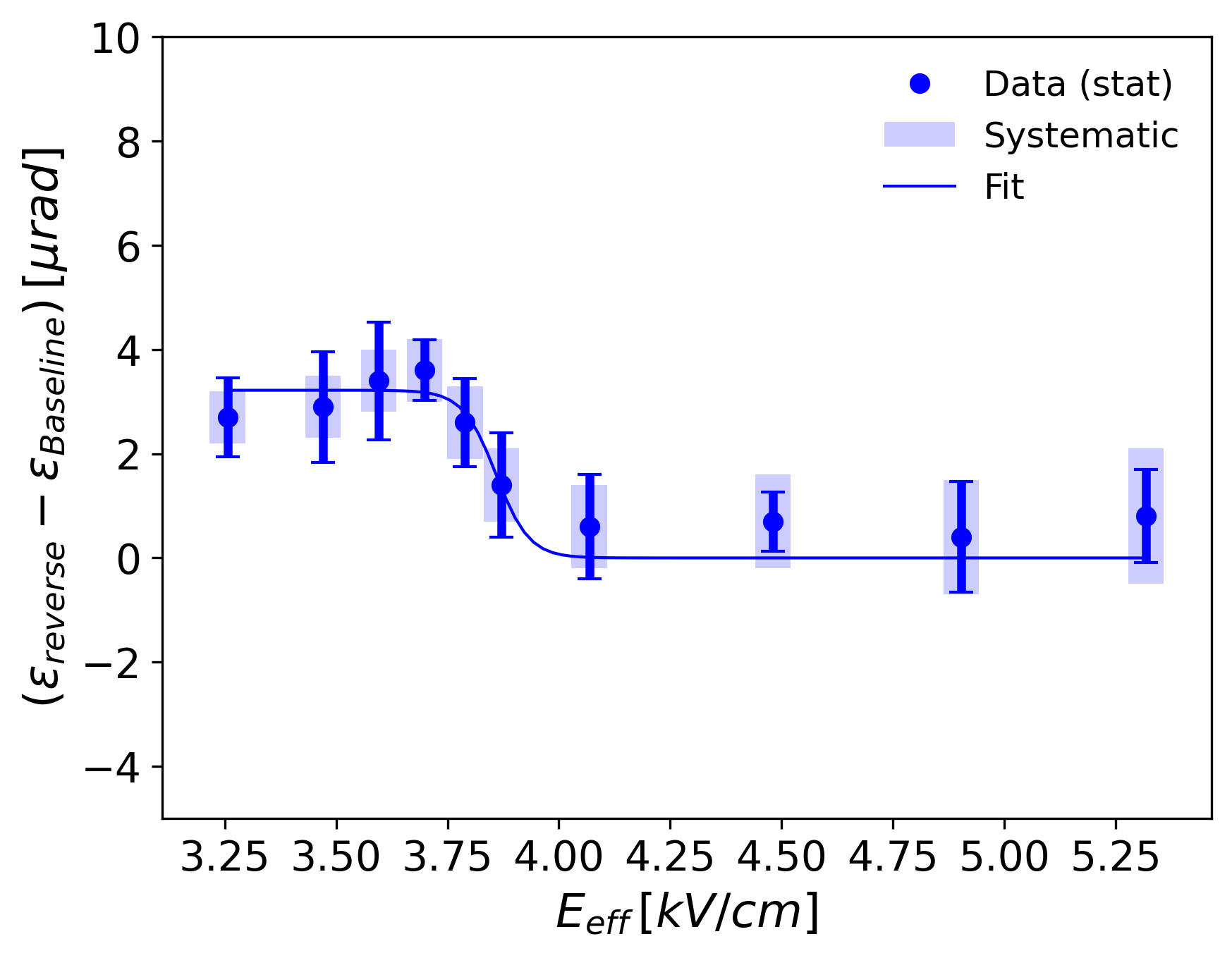}% Here is how to import EPS art
\caption{ Electric field dependence of the difference in reversed-voltage and baseline ellipticity measurements for uncoated PMMA surfaces. Shown are the statistical (blue bars) and systematic (blue boxes) uncertainties with a Fermi-type function fit.}\label{fig:pmmauncoated}
\end{figure}
%\ref{tab:table2}%
\begin{table}
\caption{Measured baseline and reversed-voltage ellipticity values and uncertainties at different effective electric fields for uncoated PMMA surfaces.}\label{tab:uncoated}
\begin{ruledtabular}
\begin{tabular}{cccccccc}
$E_{eff}$& $\epsilon_{baseline}$ & $\epsilon_{reverse}$ & $ \Delta \epsilon$ & $\delta \epsilon_{stat}$ & $ \delta \epsilon_{sys}$ & $\delta \epsilon_{tot}$ \\
\hline
3.26&$14.6\pm 0.3$  & $17.3 \pm 0.7$& 2.7 & 0.8 & 0.5 &0.9 \\
 3.47& $16.3 \pm 0.8$ & $19.2 \pm 0.7$& 2.9 &1.1 & 0.6 &1.2\\
 3.60& $18.6 \pm 0.8$& $22.0 \pm 0.8$& 3.4 & 1.1 & 0.6 &1.3\\
 3.70& $18.2 \pm 0.5$& $21.8 \pm 0.3$ & 3.6 & 0.6 & 0.6&0.8 \\
 3.79& $19.7 \pm 0.6$ & $22.3 \pm 0.6$& 2.6 & 0.8 & 0.7& 1.1\\
 3.87& $23.6 \pm 0.6$& $25.0 \pm 0.8$& 1.4 & 1.0 &0.7& 1.2\\
4.07& $20.9 \pm 0.6$& $21.5 \pm 0.8$& 0.6 & 1.0 & 0.8&1.3\\
4.48& $26.6 \pm 0.4$& $27.3 \pm 0.4$& 0.7 & 0.6 & 0.9&1.1\\
4.90& $34.7 \pm 0.8$& $35.1 \pm 0.7$& 0.4& 1.1& 1.1&1.5\\
5.32& $43.0 \pm 0.4$& $43.8 \pm 0.8$& 0.8 & 0.9&1.3&1.6\\
\end{tabular}
\end{ruledtabular}
\end{table}

A similar analysis of the difference between the reversed-voltage and baseline ellipticity measurements as a function of the effective electric field was performed for uncoated PMMA in liquid nitrogen. The results are shown in Fig.~\ref{fig:pmmauncoated}, which displays the measured data points with statistical uncertainties indicated by blue error bars and systematic uncertainties represented by blue boxes. The solid blue line shows a fit to the data using the same functional form given in Eq.~\ref{eq:fermifcn}.

The fitted values of the parameters $\epsilon_0$, $E_0$, and $\Delta E$ are
$(3.16 \pm 0.56)~\mu$rad,
$(3.86 \pm 0.06)$~kV/cm,
and $(0.04 \pm 0.06)$~kV/cm, respectively.
Figure~\ref{fig:pmmauncoated} also shows a pronounced change in the ellipticity difference, $\Delta\epsilon$, from
$(3.15 \pm 0.90)~\mu$rad at low electric fields to
$(0.62 \pm 0.90)~\mu$rad at high electric fields. This change corresponds to a $1.98\,\sigma$ shift when the total uncertainty is taken into account.

The width of the transition, defined as the electric-field interval between the points where $\Delta\epsilon$ reaches 90\% and 10\% of its maximum value, is $0.16$~kV/cm. Notably, this width is significantly smaller than the width extracted for the dTPB-coated PMMA surface ($0.51$~kV/cm). This difference may indicate a smoother effective surface for uncoated PMMA, while the larger width observed for the coated case could be associated with increased surface roughness or inhomogeneity introduced by the dTPB coating.

\par
The binding energy of ions or electrons on an uncoated PMMA surface in liquid nitrogen can be calculated using the effective field ($E_{eff}$) at which the transition occurs from a higher to a lower level. This provides information about the strength of the interaction between the surface and the ions or electrons. Assuming that the electric field at half-width is a standard value for the removal of the charged particles from the uncoated PMMA surface, the corresponding force was found to be 0.039 meV/{\AA}. Taking into account a separation distance of 3 {\AA} for the charged particles \cite{Timsina2023}, the corresponding binding energy was estimated to be 0.12 meV. This value is comparable to the binding energy per unit length measured from the dTPB-coated PMMA surface.
\section{Conclusion}
The main objective of this investigation was to extract the binding energies of ions or electrons on plastic surfaces in liquid nitrogen using the electro-optic Kerr effect. To achieve this goal, a novel method was introduced for measuring the binding energy of charges on PMMA surfaces. This method was applied to measure the binding energy of charged particles on both dTPB-coated and uncoated PMMA surfaces. The minimum magnitude of the reversed effective electric field required to induce recombination of nitrogen ions and electrons in liquid nitrogen was found to be approximately 4 kV/cm. Correspondingly, the force acting on ions or electrons was calculated to be 0.04 meV/Å, and this value was consistent for both coated and uncoated PMMA surfaces. Considering a separation distance of 3 Å (estimated from van der Waals interaction \cite{Timsina2023}), this corresponds to a binding energy of 0.12 meV for the charged particles on the PMMA surface. Notably, this experimentally measured value is lower than the estimated value assuming van der Waals interaction between ions and the PMMA surface in vacuum. While the study successfully determined the minimum magntitude for the reversed electric field required for the recombination of nitrogen ions and electrons in liquid nitrogen, it remains unclear which of the charged particles, ions or electrons, detached first from the surface and where their subsequent recombination occurred. Further experiments and in-depth analysis are required to address these questions. Additionally, the behavior of ions within liquid nitrogen, particularly whether they form snowballs akin to those observed in superfluid helium \cite{Atkins1959}, is not fully understood. The potential influence of snowballs on the binding energy of charged particles also remains uncertain. Gaining a better understanding of these phenomena could potentially reconcile the discrepancies between our experimental measurements of binding energy and the theoretical predicted values.

\begin{acknowledgements}
The authors want to thank Jiachen He and Greg Porter for useful discussions and suggestions.
This work would not have been possible without the generous funding from the Office of Nuclear Physics of the DOE Office of Science, Grant Number DE-SC0014622.
\end{acknowledgements}

\section{REFERENCES}
% \nocite{*}
\bibliographystyle{apsrev4-2}   % or unsrt, plain, revtex, etc.
\bibliography{BindingEnergies}

\end{document}